# Optimizing DDoS Detection in SDNs Through Machine Learning Models

Md. Ehsanul Haque
Computer Science and Engineering
East West University
Dhaka, Bangladesh
ehsanulhaquesohan758@gmail.com

Amran Hossain
Computer Science and Engineering
East West University
Dhaka, Bangladesh
safat3622@gmail.com

Md. Shafiqul Alam
Computer Science and Engineering
East West University, Dhaka, Bangladesh
Dhaka, Bangladesh
shafiq.14027@gmail.com

Ahsan Habib Siam
Computer Science and Engineering
East West University
Dhaka, Bangladesh
ahsan.siam365@gmail.com

Sayed Md Fazle Rabbi
Computer Science and Engineering
Ahsanullah University of Science and Technology
Dhaka, Bangladesh
sayedfazlerabbi45@gmail.com

Md. Muntasir Rahman
Computer Science and Engineering
Southeast University
Dhaka, Bangladesh
mr.abir444@gmail.com

***Abstract*—: The emergence of Software-Defined Networking (SDN) has changed the network structure by separating the control plane from the data plane. However, this innovation has also increased susceptibility to DDoS attacks. Existing detection techniques are often ineffective due to data imbalance and accuracy issues; thus, a considerable research gap exists regarding DDoS detection methods suitable for SDN contexts.**

**This research attempts to detect DDoS attacks more effectively using machine learning algorithms: RF, SVC, KNN, MLP, and XGB. For this purpose, both balanced and imbalanced datasets have been used to measure the performance of the models in terms of accuracy and AUC. Based on the analysis, we can say that RF and XGB had the perfect score, 1.0000, in the accuracy and AUC, but since XGB ended with the lowest Brier Score which indicates the highest reliability. MLP achieved an accuracy of 99.93%, SVC an accuracy of 97.65% and KNN an accuracy of 97.87%, which was the next best performers after RF and XGB. These results are consistent with the validity of SDNs as a platform for RF and XGB techniques in detecting DDoS attacks and highlights the importance of balanced datasets for improving detection against generative cyber attacks that are continually evolving.**

## I. Introduction

Software-Defined Networks (SDNs) have quickly developed because of their adaptable, programmable characteristics and separated control and data planes. [1] Yet, this adaptability leads to security flaws, specifically susceptibility to Distributed Denial of Service (DDoS) attacks that inundate network resources [2]. It is essential to detect and prevent these attacks in real-time in order to ensure the availability of the network. Machine learning methods have been successful in identifying Distributed Denial of Service (DDoS) attacks in Software Defined Networks (SDNs) by analyzing vast amounts of data and making instant choices. Random Forest, Support Vector Machine, and Deep Learning algorithms are commonly employed to enhance detection accuracy and reduce instances of false positive and negative results [3] [4]. However, as stated there are still issues like the selection of features, when the dataset is imbalanced and the computational costs are high and not resolved properly. This paper discusses some of the challenges faced and to determine the most appropriate model for solving DDoS attacks on software-defined networks using machine learning techniques [10]. The major insights and findings revealed in this research study include the following:

- Which ML model performs the best in SDN-DDoS attack detection?
- The importance of proper preprocessing to improve the accuracy of DDoS detection.
- How the techniques of data balancing improve the performance of the models.

The below sections, starting from Section 2 to Section 10, describe methodology, literature review, evaluation metrics, results and discussion, comparative analysis, justification of the results, pseudo-code for DDoS detection, conclusion, and future work. For each of these sections below, a close view has been directed towards the approach, findings, and possible future research directions in SDN-DDoS detection.

## II. Related Works

This section presents the use of machine learning models for DDoS attack detection and highlights the difference between conventional and modern approaches. Modern machine learning based methodologies can provide better network security and richer knowledge of known and unknown threat scenarios as compared to traditional mechanism.

Aktar et al. [7] proposed a deep learning model based on contractive autoencoder for DDoS in network traffic. The authors tested their model on normal and attack network traffic dataset such as CIC-IDS2017, NSL-KDD, and CIC-DDoS2019 multimedia communication, and they achieved the results accuracy between 93.41%–97.58% on CIC-DDoS2019 dataset, 96.08% on NSL-KDD, and 92.45% on CIC-IDS2017 multimedia communication respectively. As shown in Table 3 the accuracy of Aktar et al.'s is the lowest among the other methods this can be concluded that this model is in low performance among other new models.

Kumar et al. [8] developed a Long Short-Term Memory (LSTM) model for detecting DDoS threats using the CICD-DoS2019 dataset. The accuracy of the LSTM model was found highest among the rest, i.e., 98%, against 93.41% to 97.5% reported by Aktar et al. [7]. Therefore, it can be concluded that the LSTM model proposed by Kumar et al. is better than the contractive autoencoder model for DDoS detection.

Ahmed et al. [9] utilized Multilayer Perceptron (MLP) deep learning algorithm for detection of DDoS attacks in which the MLP model accuracy obtained 98.99% with dataset CTU-13 and other datasets of DDoS tools, this is higher than the accuracy 98% accomplished by Kumar et al. [8], which reflects that MLP model of Ahmed et al. is high superiority accuracy rather than LSTM model.

Garba et al. [11] proposed a real-time DDoS detection and mitigation framework using machine learning models including Decision Trees, Support Vector Machines (SVM), and Logistic Regression. The Decision Tree algorithm in this research obtained an accuracy of 99%. It is higher than the accuracy of 98.99% attained by Ahmed et al.'s [9], this indicates that the Decision Tree model of Garba et al. has better accuracy in DDoS attack detection than Ahmed et al.'s [9].

Nalayini et al. [12] evaluated eight machine learning algorithms for DDoS detection using the CIC-IDS2017 dataset and obtained an accuracy of 99.885%. This result is higher than the result reported by Garba et al. [11], where they reported 99%. Thus, it can be concluded that the Random Forest model from Nalayini et al.'s [12]study is more accurate in detecting DDoS attacks than the Decision Tree model.

Kumari et al. [13] propose a mathematical model to detect DDoS attack using Logistic Regression and Naive Bayes. The CAIDA 2007 Dataset and the Weka data mining tool were used for their experiment. The result showed that Logistic Regression in accuracy from 99% to 100%, was the highest over other methods aforementioned before. The accuracy gotten by logistic regression from Kumari et al.'s is higher than Rudro at el.'s [14] which was 99.4%which shows the Logistic Regression has the best performance values for detection of DDoS attacks over others as well.

Rudro et al. [14] also utilized the SDN dataset to evaluate machine learning approaches for DDoS detection and generated accuracy of 99.4% from the Random Forest algorithm. While it is still lower than Nalayini et al.'s finding [12] with 99.885%, but the Random Forest model constructed by Rudro et al. is already a very high accuracy in DDoS detection.

Sk Dash et al. [15] present the detection of DDoS attacks within IoT networks using the NSL-KDD dataset. They discuss two approaches: one with Principal Component Analysis, and one without, each with the inclusion of steps for preprocessing using robust scaling and encoding. The results show a huge improvement in efficiency, especially by the Random Forest and KNN classifiers, to 99.87% and 99.14%, respectively. Indeed, this work also gives special attention to the different techniques of preprocessing with a view to enhancing security against DDoS attacks in IoT systems.

## III. Methodology

The methodology is the step-by-step systematic approach that describes conducting research or a study by which methods, procedures, and techniques are used in carrying out the collection of data, analysis, and presentation of conclusions [5]. In this paper, the methodology is intended to be employed toward the detection of DDoS attacks in Software-Defined Networking through machine learning.

### A. Dataset collection

We obtained a DDoS attack dataset from Kaggle [6], which contains 104,345 records with 22 features. This dataset is for DDoS attack detection. Each record in the dataset includes different network traffic metrics for analysis and pre-detection of DDoS attacks. Table I provides the summmary of the dataset.

TABLE I
Label Distribution in Training Sets

| Dataset Type | Label | Count | Description |
|---|---|---|---|
| **Total Data** | 0 | 63,561 | Benign |
| | 1 | 40,784 | DDoS |
| | | **104,345** | **Total** |

### B. Preprocessing

We used the SDN DDoS dataset from Kaggle to build and test Software-Defined Networks (SDN) Distributed Denial of Service (DDoS) attack detection models. We followed a methodical process that included several stages for dataset preprocessing, model training and evaluation.

**Dataset Examination and Preparation:** The total number of entries and columns in the SDN DDoS dataset are determined and their data type is described in details. Out of the entire data set it is split into the training data set, which is 80% and the test data set which constitute 20%.

**Data Cleaning and Encoding:** Missing values were filled using appropiate methods. Also, to make the categorical features compatible with the machine learning algorithms we have converted them into numerical form using encoding techniques.

**Class Imbalance Handling:** To address the problem of class imbalance, the minority attack classes in the training dataset were over-sampled using SMOTE so that the datasets used for training and testing were balanced but the testing dataset remained pure real-world representation.

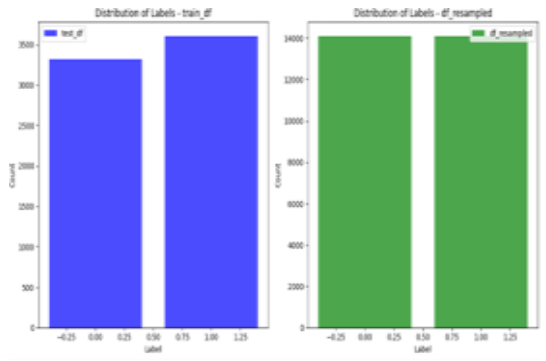

Fig. 1. Before and After Data Balancing

**Outlier Removal and Standardization**

We removes outliers using the Local Outlier Factor (LOF) and then standardizes large values to ensure consistency across features.

**Model Implementation:**

We have applied different machine learning model for DDoS detection such as Random Forest, Support Vector Machine (SVM), Multi-Layer Perceptron (MLP), K-Nearest Neighbors (KNN).

**Model Evaluation:** We tested the model performance using accuracy, precision, recall, F1-score and AUC-ROC metrics. The test data was preprocessed in the same manner as training data except class balancing to get the right estimate in live production.

**Validation and Tuning:** Further, to make models more trustable, we performed hyperparameter tuning and cross-validation on the models to ensure that the models were no only accurate, reliable but also still had great generalization capabilities with other data portions.

This integrated approach was useful in the establishment of an effective DDoS detection technique having high levels of identification of malicious behaviors in the SDN environment and optimizing multiple machine learning techniques for enhanced performance.

Below figure 2 represent the proposed working flow

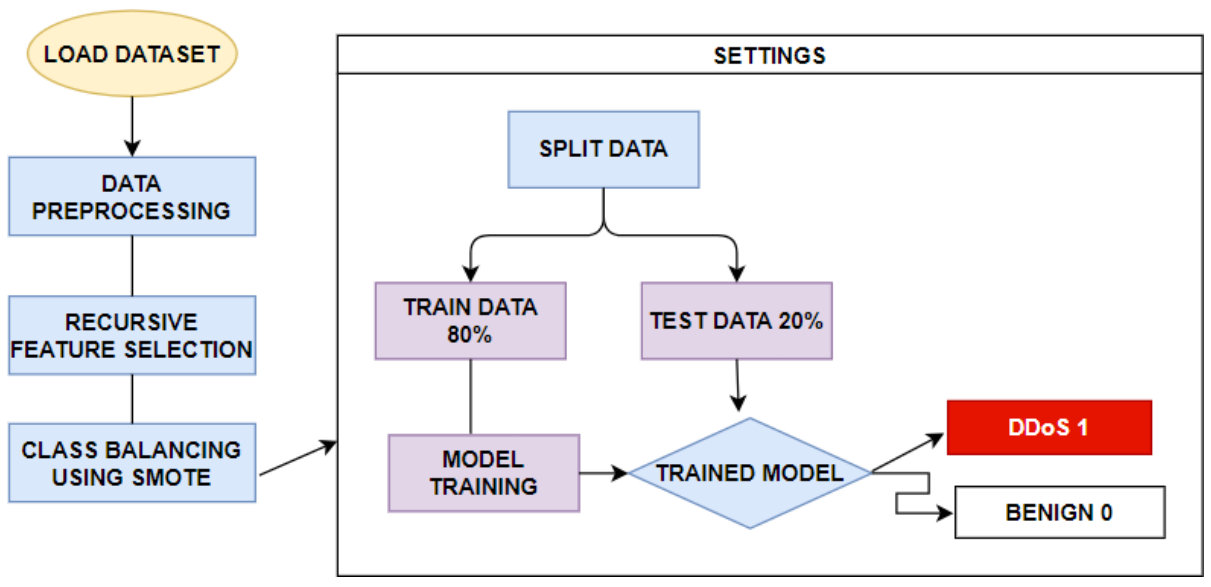

Fig. 2. Workflow Diagram

### C. *Evaluation Metrics*

The commonly used models performance metrics are accuracy, precision, recall, F1 score, and AUC-ROC and confusion matrix described in this section.

$$\text{Accuracy} = \frac{TP + TN}{TP + TN + FP + FN}$$

$$\text{Precision} = \frac{TP}{TP + FP}$$

$$\text{Recall} = \frac{TP}{TP + FN}$$

$$\text{F1-score} = \frac{2 \cdot \text{Precision} \cdot \text{Recall}}{\text{Precision} + \text{Recall}}$$

Confusion Matrix: A table summarizing the performance of a classification model, showing true positives (TP), true negatives (TN), false positives (FP) and false negatives (FN).

AUC: Measures a model's ability to distinguish between classes. It represents the area under the ROC curve, with higher values indicating better performance.

## IV. Results and Discussions

Here, the results obtained from the different machine learning models trained and tested with both imbalanced and balanced datasets are provided. The evaluation concerns performance indicators including accuracy and Area Under the Curve, or AUC, scores. Considering the classification algorithms: Random Forest (RF) and Support Vector Classifier (SVC), k-Nearest Neighbors (KNN), Naive Bayes (NLP, XGBoost (XGB). The comparison also reveals how the application of dataset balancing affect the performance of these models for DDoS attack detection in Software-Defined Networks (SDNs).

**Performance on Imbalanced Dataset**

**Training**

TABLE II
Model Training Performance Summary without Balancing

| Metric | RF | SVC | KNN | MLP | XGB |
|---|---|---|---|---|---|
| Training Accuracy | 1.0000 | 0.9799 | 0.9882 | 0.9992 | 1.0000 |
| Mean CV Accuracy | 0.9998 | 0.9780 | 0.9814 | 0.9981 | 0.9998 |

In the table II, the training and cross validation results are shown with respect to an imbalanced training dataset.The training accuracy of each model represents the classifier's capability of predicting the training data correctly; thus, Random Forest (RF) and XGBoost (XGB) record the maximum training accuracy of 1.0000. The Support Vector Classifier (SVC) and K-Nearest Neighbors (KNN) algorithms also perform quite high with their training accuracy reaching 0.9799 and 0.9882 respectively.

The mean cross-validation accuracy shows the capacity for generalization of the corresponding models, where Random

Forest and XGBoost also lead with mean CV accurracies of 0.9998. The mean CV accuracy of SVC and KNN is 0.9780 and 0.9814 respectively, also the Multi-layer Perceptron (MLP) is good with the average CV accuracy of 0.9981. It indicates the efficiency and reliability of the training process for all models with respect to the task objective.

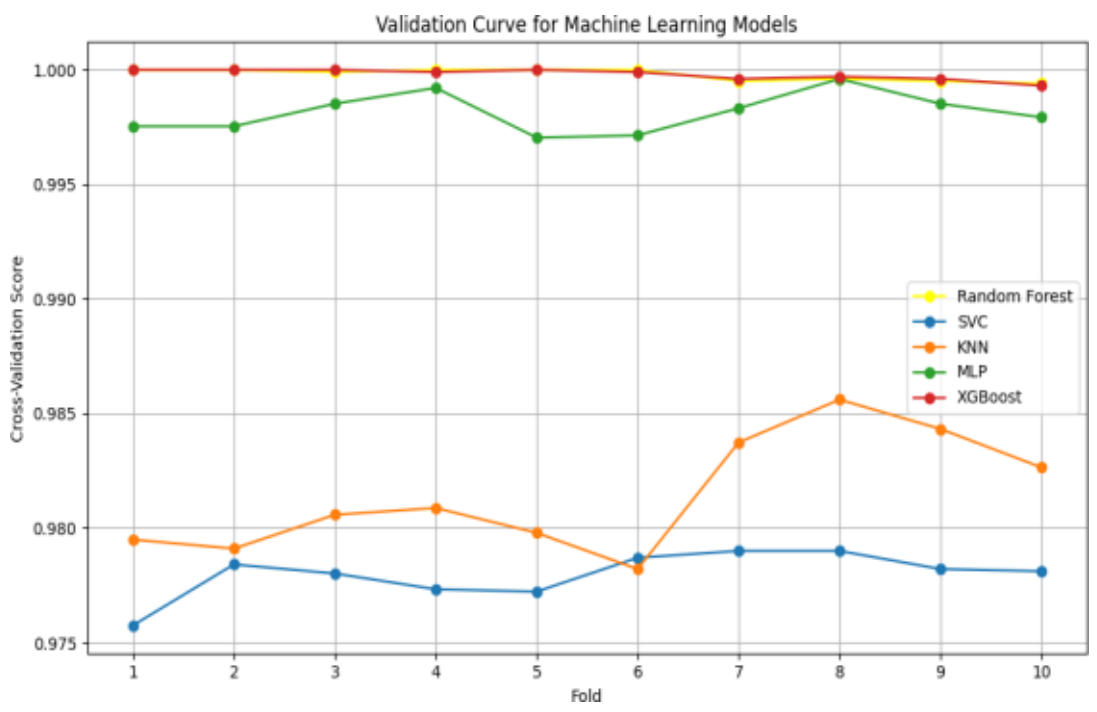

Fig. 3. Validation curve

### Testing

TABLE III
TEST ACCURACY AND AUC FOR DIFFERENT CLASSIFIERS

| Model | RF | SVC | KNN | MLP | XGB |
|---|---|---|---|---|---|
| **Accuracy** | 1.0000 | 0.9747 | 0.9778 | 0.9987 | 1.0000 |
| **AUC** | 1.0000 | 0.9971 | 0.9937 | 1.0000 | 1.0000 |

The table V presents testing results for the trained models in the imbalanced dataset proves all models have a high level of accuracy.The Random Forest and XGBoost Classifier shows the highest level of accuracy and area under the curve of 1.0000.Also, Multi-layer Perceptron testing accuracy of 0.9987 and AUC 1.0000. The Support Vector Classifier was slightly less accurate at 0.9747 along with AUC at 0.9971 while the KNN classifier was slightly lower at 0.9778 with the AUC at 0,9937, which denotes high predictive proficiency.

## Performance on Balanced Dataset

### Training

TABLE IV
MODEL TRAINING PERFORMANCE SUMMARY WITH BALANCING

| Metric | RF | SVC | KNN | MLP | XGB |
|---|---|---|---|---|---|
| Training Accuracy | 1.0000 | 0.9784 | 0.9851 | 0.9998 | 1.0000 |
| Mean CV Accuracy | 1.0000 | 0.9758 | 0.9762 | 0.9986 | 1.0000 |

Table IV shows the training accuracy on balanced dataset. The Random Forest and XGBoost achieved highest training accuracy as an balance dataset is 1.00. Also, the cross-validation accuracy is exceptional for these two models as it achieved perfect cross validation accuracy 1. Others models also performs well but bit lower than the top performing models like RF and XGBoost . Above all, SVC achieved the lowest training and cross validation accuracy in balanced dataset.

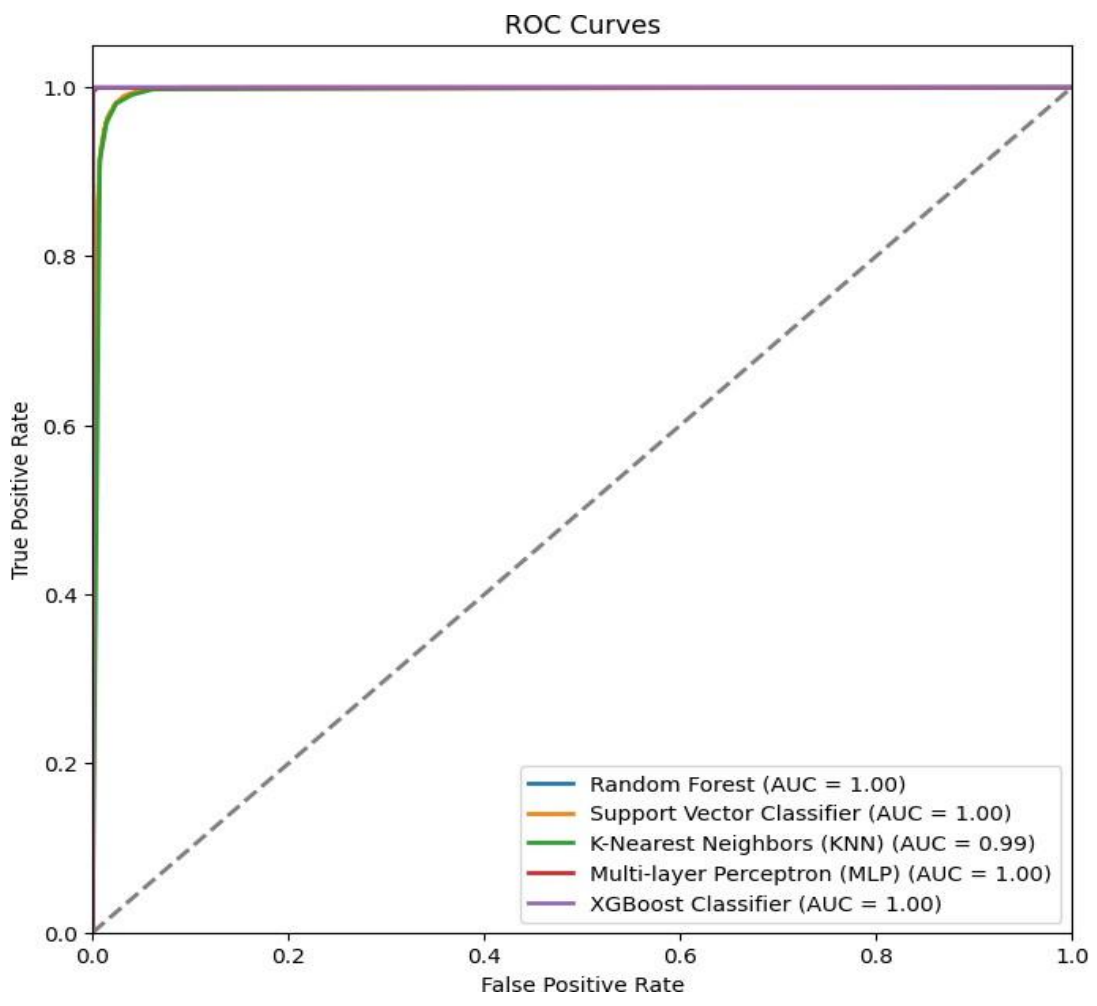

Fig. 4. ROC CURVE

The validation curve on figure 6 shows performance of various machine learning models was evaluated through cross-validation. Random Forest and XGBoost achieved perfect accuracy (1.0000), while Support Vector Classifier had a mean accuracy of 0.9758.

### Testing

TABLE V
MODEL TEST ACCURACY AND AUC

| Model | RF | SVC | KNN | MLP | XGB |
|---|---|---|---|---|---|
| **Test Accuracy** | 1.0000 | 0.9765 | 0.9787 | 0.9993 | 1.0000 |
| **AUC** | 1.0000 | 0.9972 | 0.9953 | 1.0000 | 1.0000 |

Table V displays the test accuracy and AUC values for different machine learning models that are used in this dataset to detect DDoS. The Random Forest and XGBoost performs exceptionally as they achieves perfect accuracy for both evaluation metrics. Also, MLP achieves near perfect accuracy with an AUC 1.00. On the other hand, SVC and KNN show slightly lower performance compared to other models but the accuracies and AUC's are acceptable and above 97.65%. Overall. These results show slightly improved performance than imbalance dataset.
Below figure 7 represent the ROC curve and Confusion matrix of all models.

The figure 8 shows that both XGBoost and Random Forest models have no false positives (FP) or false negatives (FN), demonstrating exceptional classification accuracy. In contrast, the Support Vector Classifier (SVC) shows a higher rate of false positives but now much lower than imbalance dataset, misclassifying some negative instances as positive, highlighting its relative performance issues compared to the other models.

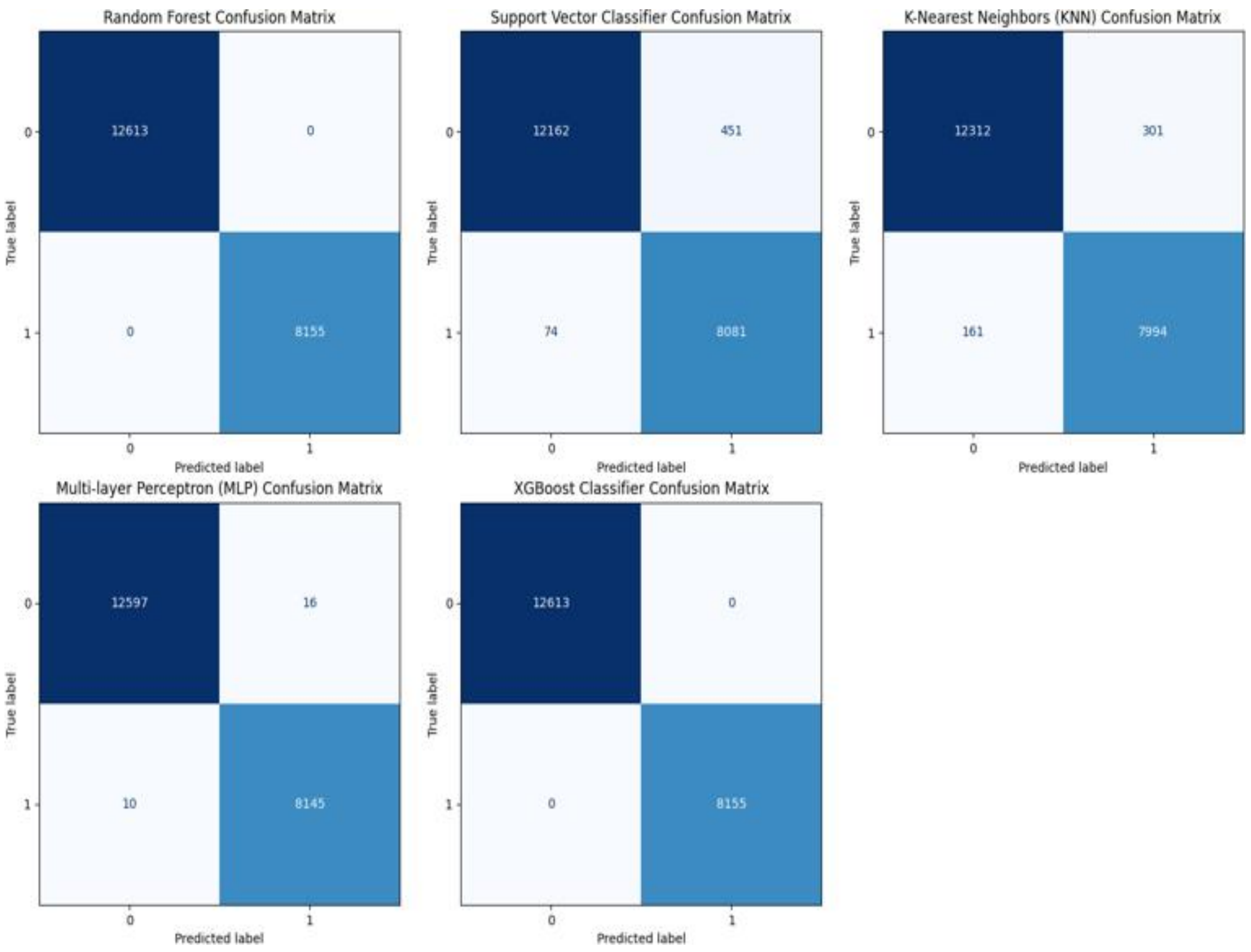

Fig. 5. Confusion Matrix

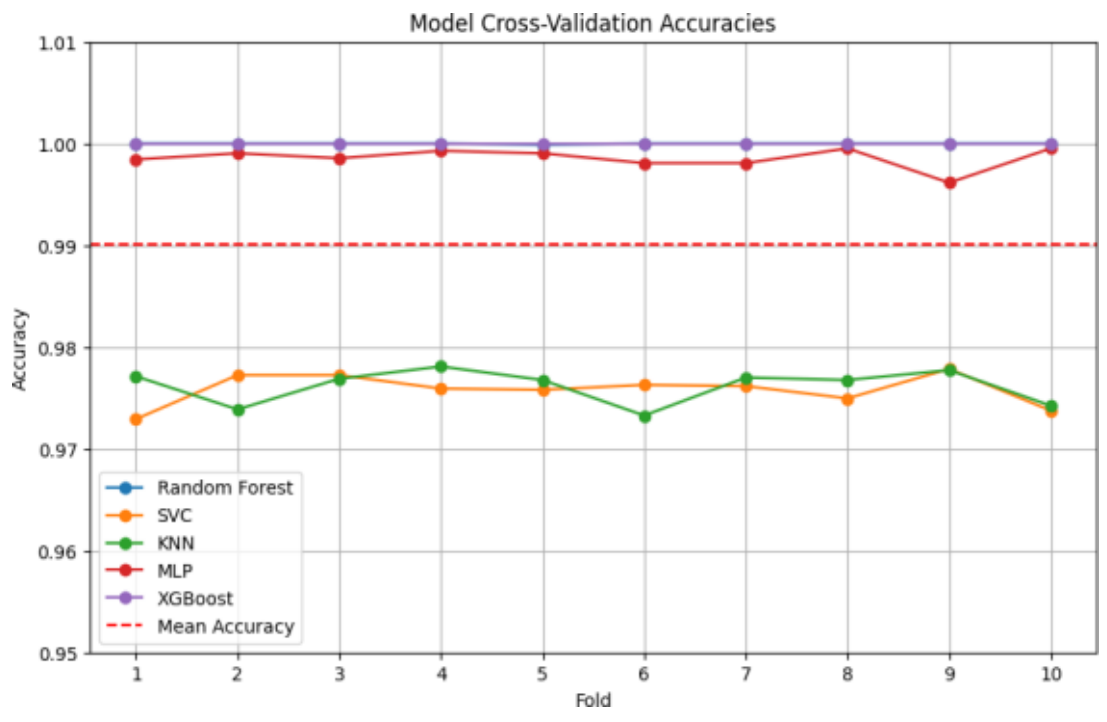

Fig. 6. Validation Curve

## V. Comparative Analysis

TABLE VI
Comparative Analysis of DDoS Detection Methods

| Study | Method | Dataset | Accuracy |
|---|---|---|---|
| [13] | Logistic Regression | CAIDA 2007 | Above 99% |
| [12] | Random Forest | CIC-IDS2017 | 99.885% |
| [8] | LSTM | CICDDoS2019 | Up to 98% |
| [15] | RF | NSL-KDD | 99.87% |
| [11] | Decision Tree (DT) | SDN | 99% |
| [14] | Random Forest | SDN | 99.4% |
| **Proposed Work** | **XGB** | **SDN** | **100%** |

Table VI presents a comparative analysis of the accuracies of different DDoS detection techniques with the existing literature, including our proposed Extreme Gradient Boosting (XGB) model, which achieved an exceptional 100% accuracy in both balance and imbalance dataset. Also, out study surpassed most of the existing literature in a efficient way.

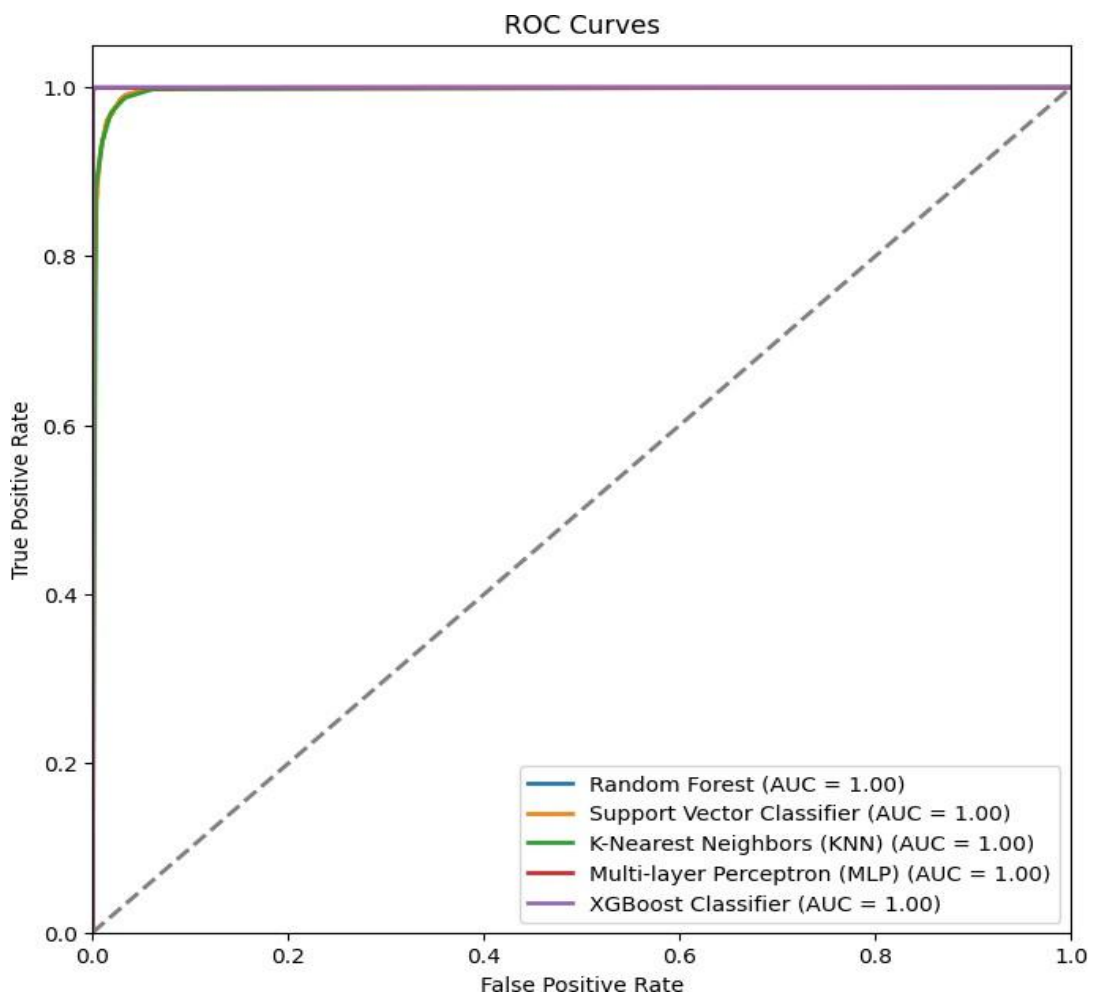

Fig. 7. ROC Curve

## VI. Justification of results

The models Random Forest (RF) and XGBoost (XGB) achieved perfect accuracy of 1.0, while Support Vector Classifier, K-Nearest Neighbors, and Multi-layer Perceptron performed exceptionally well, with accuracies between 97% and 99%. This raises questions about the reliability of these results: Are they truly indicative of model performance, or do they suggest potential overfitting? To investigate this, metrics such as Cohen's Kappa, Matthews Correlation Coefficient (MCC), and Brier Score were used. These metrics assess the agreement between predicted and actual outcomes, the balance of predictions, and the accuracy of probabilistic predictions. The evaluation results are summarized in Table VII.

TABLE VII
Model Evaluation Metrics

| Model | Cohen's Kappa | MCC | Brier Score |
|---|---|---|---|
| Random Forest | 1.0000 | 1.0000 | 0.0001 |
| Support Vector Classifier | 0.9474 | 0.9481 | 0.0177 |
| K-Nearest Neighbors (KNN) | 0.9535 | 0.9536 | 0.0176 |
| Multi-layer Perceptron (MLP) | 0.9981 | 0.9981 | 0.0009 |
| XGBoost Classifier | 1.0000 | 1.0000 | 0.0000 |

Table VII presents evaluation metrics for machine learning models used to detect DDoS attacks in Software-Defined Networks (SDNs). Both Random Forest and XGBoost Classifier achieved perfect Cohen's Kappa and Matthews Correlation Coefficient (MCC) scores, indicating exceptional accuracy and strong agreement with actual attack classifications. Their low Brier Scores further demonstrated minimal prediction errors.

Support Vector Classifier and K-Nearest Neighbors also performed well, with Kappa and MCC scores above 0.94, indicating reliable detection capabilities, although slightly lower than the top models. The Multi-layer Perceptron showed high Kappa and MCC values with a low Brier Score, confirming its effectiveness in DDoS detection.

In summary, while all models performed well, Random Forest and XGBoost were particularly reliable, with XGBoost emerging as a robust option for practical applications in SDN environments.

## VII. Pseudocode of DDoS Detection

**Algorithm 1** DDoS Detection Using Balanced and Imbalanced Datasets

**Step 1: Load Dataset**
**Step 2: Data Exploration and Cleaning**
Explore dataset, handle missing values, remove irrelevant features, encode categorical variables, normalize numerical features.
**Step 3: Class Distribution**
**Step 4: Class Imbalance Handling**
Apply oversampling (e.g., SMOTE).
**Step 5: Feature Selection**
Select important features based on model-based analysis.
**Step 6: Data Splitting**
**Step 7: Model Training**
Train models (RF, SVM, KNN, MLP, XGBoost) on both imbalanced and balanced datasets.
Tune hyperparameters using cross-validation.
**Step 8: Evaluation**
Evaluate models (Accuracy, Precision, Recall, F1-score, AUC).
**Step 10: Result Summary**
Summarize findings.

## VIII. Conclusion

This paper evaluates several machine learning algorithms to detect DDoS attacks in Software Defined Networks, such as Random Forest, Support Vector Classifier (SVC), K-Nearest Neighbors (KNN), Multi-layer Perceptron (MLP), and XGBoost. The findings were such that for DDoS event identification accuracy, Random Forest and XGBoost scored 1.0000 perfect accuracy and showed their outstanding performance in the correct detection of DDoS events. These models also scored good scores for Cohen's Kappa and Matthews Correlation Coefficient, which means they are highly in agreement with the actual classification of attacks. On the other hand, although SVC, KNN, and MLP have an excellent performance with an accuracy between 97% to 99%, it is not as high as the best-performance RF and XGB. Results underline the need for effective detection systems against changing DDoS threats, with XGBoost emerging as particularly promising for real-world use within SDNs.

## IX. Future Work

Future work should be directed with larger and more diverse datasets, including data from attacks in real-time to train and test the models. In addition, studies of hybrid machine learning and ensemble methods could further enhance detection capability. Further studies with advanced sampling methods to balance data as well as feature engineering will be required for better intrusion detection systems. Finally, validation of the models in dynamic SDN environments will be critical to ensure their effectiveness against new cyber adversaries.

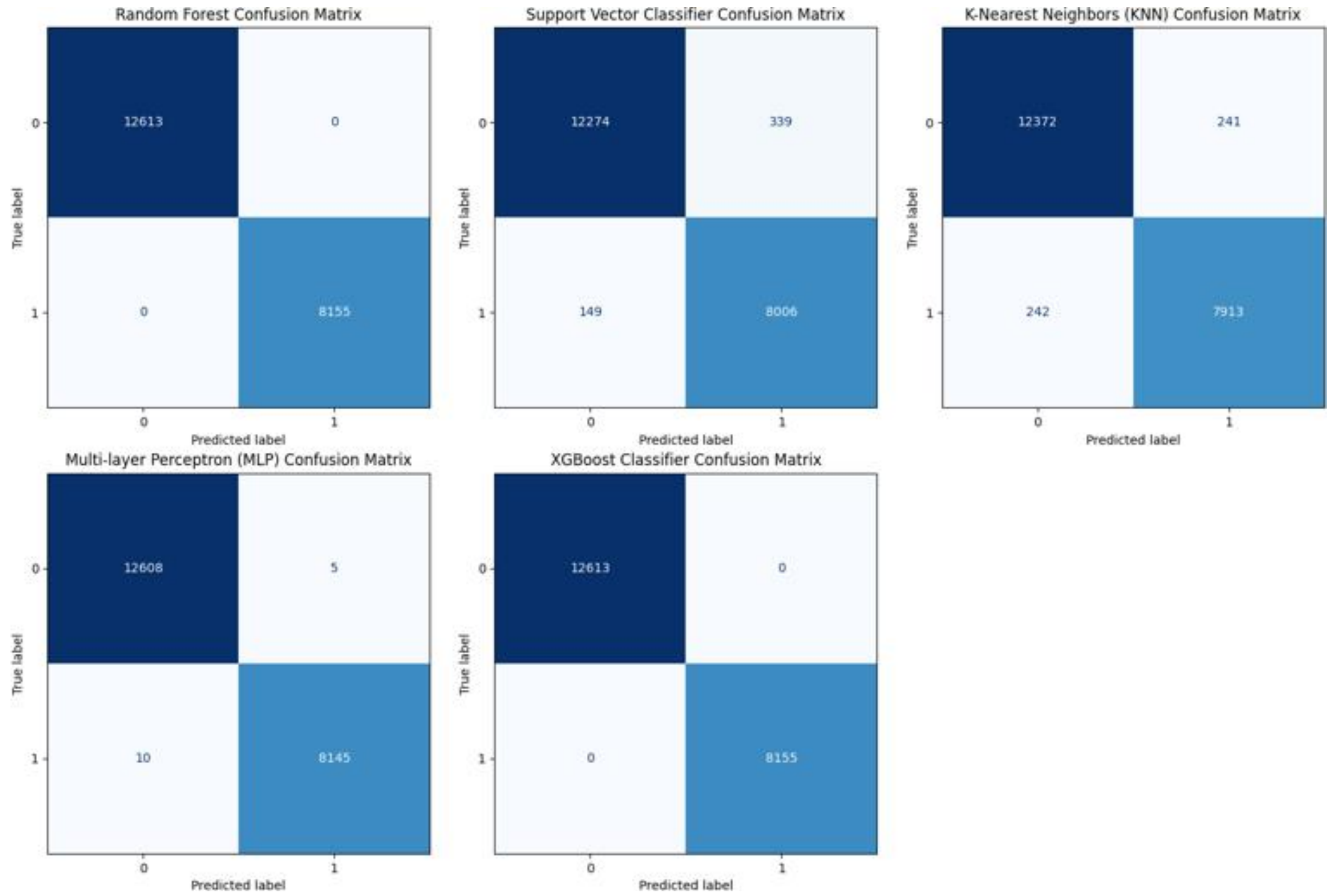

Fig. 8. Confusion matrix